\documentclass[12pt]{iopart}
\begin{document}
\input{epsf.tex}
\epsfverbosetrue

\title{Nonlinear optical beams carrying phase dislocations}

\author{Anton Desyatnikov \dag\ \footnote[3]{desya@uni-muenster.de}, Cornelia
Denz \dag, and Yuri Kivshar \ddag}

\address{\dag\ Nonlinear Photonics Group, Institute of Applied
Physics, Westf\"{a}lische Wilhelms-Universit\"{a}t M\"{u}nster,
D-48149 M\"{u}nster, Germany}

\address{\ddag\ Nonlinear Physics Group, Research School of Physical Sciences and Engineering,
Australian National University, Canberra ACT 0200, Australia}

\begin{abstract}
We describe different types of self-trapped optical beams carrying
phase dislocations, including {\em vortex solitons} and ring-like
{\em soliton clusters}. We demonstrate numerically how to create
such nonlinear singular beams by the interaction of several
fundamental optical solitons. Mutual trapping of several solitons
can be regarded as a synthesis of {\em `soliton molecules'}, and
it corresponds to a transfer of an initial orbital angular
momentum of a system of solitons to a spin momentum of an optical
vortex.
\end{abstract}


\submitto{\JOA: Singular Optics Special Issue}

\section{Introduction}

Phase dislocations carried by the wavefront of a light beam are
associated with the zero-intensity points where the light
intensity vanishes. The phase of the wave is twisted around such
points creating a structure associated with {\em an optical
vortex}. Optical beams with phase dislocations play an important
role in linear singular optics~\cite{review}.

In self-focusing nonlinear media, an intense finite-extent laser
beam becomes localized due to the self-trapping mechanism which
can compensate for the beam diffraction. The nonlinear self-action
of light may result in the formation of stationary structures with
both intensity and phase remaining unchanged along the propagation
direction. Such self-trapped stationary structures of light beams
are termed {\em spatial optical solitons}~\cite{book}. When such
solitons have phase singularities, they determine the internal
structure of the beam; they can be stabilized by the light
self-action generating nonlinear self-trapped optical beams
carrying phase dislocations. Examples of such beams include {\em
vortex solitons}~\cite{krug,Skryabin,vort-opn} with point screw
dislocations, {\em multipole vector solitons}~\cite{multipole}
with $\pi$-edge dislocations, and more complicated {\em
`necklace'}-type beams~\cite{necklace,neckrot,neckring} and {\em
soliton clusters}~\cite{cluster} with a combination of a screw
dislocation at the beam center and, generally, $\vartheta$-edge
dislocations, where $\vartheta$ is the phase jump between
neighboring peaks in the intensity distribution~\cite{cluster}.

The fundamental optical solitons show a fascinating combination of
the properties of classical wave-packets together with a number of
particle-like properties demonstrated in their elastic and
inelastic interactions and mutual scattering, when each of the
solitons preserves its identity. Moreover, the coherent
interaction between the solitons depends strongly on a relative
phase what provides an additional degree of freedom to control the
interaction. We may draw an analogy between spatial soliton and
the `atom of light', and then the soliton collision and
interaction can be treated in terms of the effective forces acting
between these effective `atoms'. Following this concept, the
higher-order multi-hump optical beams can be regarded as bound
states of `atoms' trapped by a common potential induced in a
nonlinear medium. A balance of the interaction forces acting
between the solitons is the necessary condition for the formation
of the soliton `clusters' or `molecules of light'.

In this paper we investigate the excitation of higher-order beams,
including optical vortices and soliton clusters, through the
inelastic soliton scattering and mutual trapping of initially
well-separated fundamental solitons, the effect resembling a
synthesis of `soliton molecules'. In addition, we propose the
application of this effect in the context of `soliton algebra'
\cite{algebra} regarding the fundamental spatial solitons as the
information carriers, and the transformation of an optical pattern
induced by the soliton interaction as all-optical soliton
switching.

\section{Optical vortex solitons and soliton clusters}

Optical vortices were introduced as the first example of a
stationary light beam with the phase twisted around its core; the
twist is proportional to $2\pi$ with integer $m$, the so-called
topological charge of the phase dislocation~\cite{krug}. The
physical model analyzing the evolution of the slowly varying field
envelope $E$ is described by the nonlinear Schr\"odinger equation,
\begin{equation}
\label{eq1}\ i\frac{\partial E}{\partial z}+\Delta _{\perp
}E+f(I)E=0,
\end{equation}
where $\Delta _{\perp}$ is the transverse Laplacian, and $z$ is
the propagation distance measured in the units of the diffraction
length. Function $f(I)$ describes the nonlinear properties of an
optical medium, and it is assumed to depend on the total beam
intensity, $I=\left|E\right|^{2}$. The simplest spatially
localized solution of equation \eref{eq1} carrying a phase
dislocation, i.e. the vortex soliton, can be written in the form:
$E(r,\theta,z) = A(r)\e^{i m\theta + i\beta z}$, where $A(r)$ and
$\beta$ are the beam amplitude and propagation constant,
respectively, while $r$ and $\theta$ are the polar coordinates in
the transverse plane.

In self-focusing nonlinear media, such vortex beams are the
subject of the azimuthal modulational instability which result in
splitting of the doughnut ring-like structure into a certain
number of the fundamental solitons. The number of splitters and
their dynamics are determined by the topological charge of the
phase dislocation corresponding to the beam angular momentum (see,
e.g.,~\cite{Skryabin} and references therein).

\begin{figure}
\begin{center}
\epsfbox{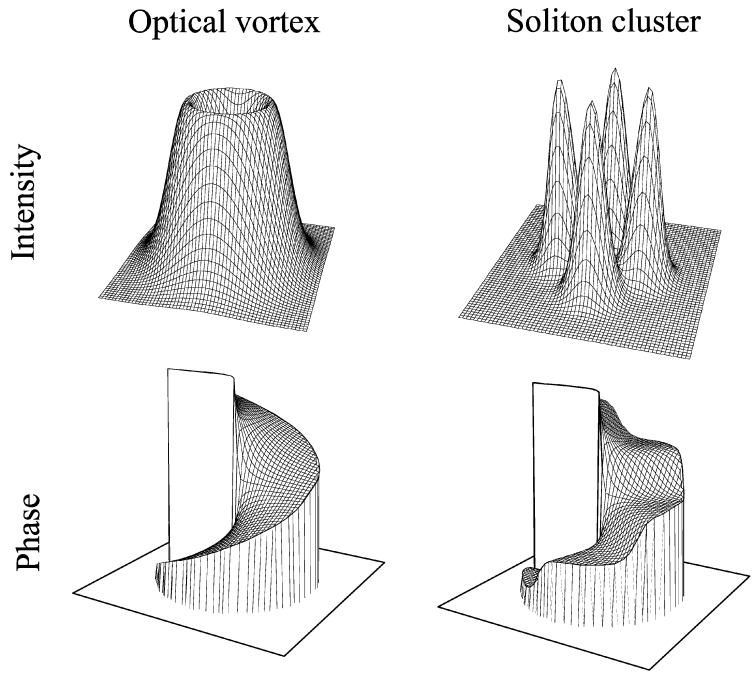}
\end{center}
\caption{\label{fig1} Intensity and phase distributions for
optical vortex soliton and soliton cluster composed of four
fundamental beams. Note that in terms of the azimuthal coordinate
$\theta =\tan^{-1}(y/x)$, the vortex phase is given as a {\em
linear function} $m\theta$ with integer $m$, while the
staircase-like phase of the cluster is a {\em nonlinear} phase
dislocation.}
\end{figure}

The simplest higher-order scalar stationary solutions found for
the model \eref{eq1} are a family of the radially symmetric
solitons which includes radial modes with nodes in form of
concentric rings. The important characteristic of these states is
the {\em soliton spin} determined as a ratio of two conserved
quantities, the beam angular momentum and the beam power. For the
vortex soliton, the spin coincides with the topological charge $m$
of the phase dislocation carried by the vortex. Because of the
condition of the field univocally, the topological charge $m$ is
quantized and has integer value. The fundamental spatial solitons
and their higher-order radial states have zero spin.

Novel types of the higher-order self-trapped optical beams can be
introduced as {\em azimuthally modulated} self-trapped structures
in the form of the so-called `necklace'
beams~\cite{necklace,neckrot}. However, it was found that a
combination of the edge-phase dislocation with $\pi$-out-of-phase
neighbor peaks cannot produce a stationary state, and the
structure becomes slowly expanding~\cite{necklace}. Such a
stabilization is indeed possible for a more complicated system
including the attraction between several incoherent
beams~\cite{multipole,neckring}. Another approach to this problem
is to combine the screw dislocation in the origin of a ring-shaped
beam with the edge dislocation within the necklace~\cite{neckrot}.
The screw dislocation introduces a centrifugal force to the ring,
being also responsible for spiralling and mutual repulsion of the
solitons in the case of vortex break-up~\cite{Skryabin}, and the
edge dislocations prevents noise-induced instability break-up of
the ring. Because of a nonzero angular momentum, the whole
structure rotates with its propagation. As a result, the
stabilization of the ring-shaped multihump beams requires complex
phase distribution characterized by {\em fractional} value of the
soliton spin~\cite{neckrot,neckring}.

A phase distribution required for the formation of
quasi-stationary higher-order self-trapped optical beams was found
in references~\cite{cluster} where the concept of {\em soliton cluster}
was introduced. In this approach, the azimuthally modulated beam
is regarded as a bound state of the interacting fundamental
solitons. Because of phase-sensitive interaction, the requirement
of the balance of the interaction forces between the solitons
determines the beam phase in the form of a {\em staircase-like}
screw dislocation. \Fref{fig1} compares the vortex phase
dislocation (left column) with the phase of a four-soliton
cluster, having well defined $\pi/2$ steps between the soliton
positions (right column). It was found~\cite{cluster} that a {\em
radially stable} dynamical bound state is formed if these
phase-jumps satisfy the condition $\vartheta = 2\pi m/N$, with
$N\geq 4m$ being the number of solitons in the ring.

Stability of the soliton clusters has been tested numerically for
different nonlinear media, including cubic saturating, competing
cubic self-focusing and quintic self-defocusing, and competing
quadratic and cubic self-defocusing nonlinearities \cite{cluster}.
The idea has been also extended to higher dimensions, covering the
case of the spatio-temporal vortex solitons and light bullets. The
common outcome of these studies is the confirmed robustness of the
soliton clusters to random noise and strong radial perturbations.
In the latter case, the pulsating states viewed as the radial
excitations of `soliton molecule' have been observed.
Nevertheless, soliton clusters are the subject of modulational
instability, and they are unstable with respect to azimuthal
perturbations. The remarkable feature of this instability is that
the number of the fundamental solitons flying off the main ring
after the splitting is determined mainly by the topological charge
$m$ instead of the initial number $N$ of solitons, similar to the
vortex solitons. For what follows, we stress the fact that the
conservation of the angular momentum of an optical beam in an
isotropic medium determines the dynamics of splitters after
break-up, so that the initial `spin' angular momentum of vortex or
cluster can be viewed as being {\em transformed} into the orbital
momentum of the spiralling splitters \cite{Skryabin}.

\section{Soliton molecules}

Recent progress in both theoretical and experimental studies of
the higher-order optical spatial solitons brought the soliton
community to the gates of the direct search for all-optical
soliton-based switching schemes, when the initial data carried by
the light distribution on the front-face of the nonlinear medium
can be processed, in a predictable and controllable way, by
employing the light self-action effects. One of the examples of
such an approach is the recently proposed concept of the `soliton
algebra' \cite{algebra}, based on the instability-induced break-up
of optical vortices to the controllable number of the fundamental
solitons. This idea also represents the example of a nontrivial
approach to the soliton instability, when the symmetry-breaking
instability, usually regarded as a serious disadvantage in using
spatial solitons, is employed as {\em a key physical mechanism}
for all-optical soliton switching from a given initial state
(optical vortex) to the known final state defined by a certain
number of fundamental solitons. This approach can be generalized
to a broad variety of scalar and vector higher-order {\em
metastable} solitons.

The symmetry-breaking soliton instability may serve as a physical
mechanism for all-optical switching with only one disadvantage --
it is a one-way process describing the transition from an initial
higher-order state (a soliton molecule or cluster) to a number of
simple stable states (dipole-mode and fundamental solitons).
Below, we propose the opposite process, viewed as the excitation
or `synthesis' of higher-order states from predefined number of
initially separated solitons, or `atoms of light', in a nonlinear
bulk medium. Indeed, introducing molecules of light would not be
self-consistent without the possibility of mutual trapping of the
free atoms or molecule synthesis.

\begin{figure}
\begin{center}
\epsfbox{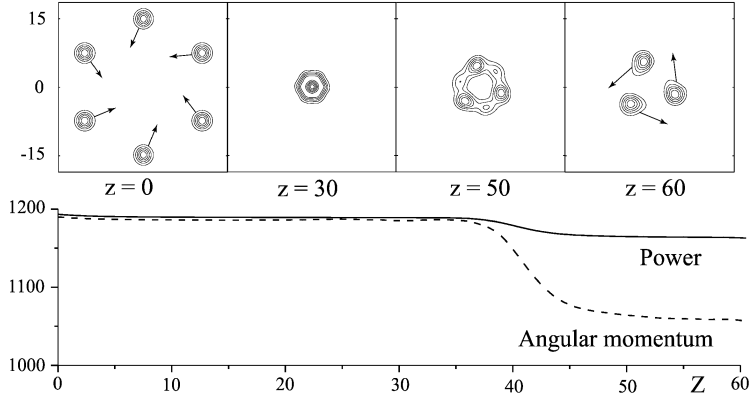}
\end{center}
\caption{\label{fig63} Switching from $N=6$ to $N=3$ solitons.
After reaching the minimum radius at $z=30$, the {\em dynamically
unstable} three-lobe cluster is formed with its subsequent
break-up into three fundamental solitons flying away. Note that
symmetry-breaking instability is accompanied by power and angular
momentum loses due to nonsolitonic radiation emission.}
\end{figure}

\begin{figure}
\begin{center}
\epsfbox{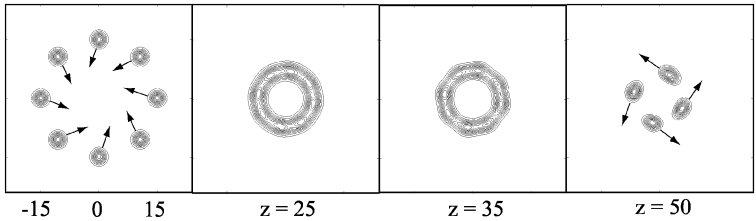}
\end{center}
\caption{\label{fig84} Switching from $N=8$ to $N=4$ solitons.
After reaching the minimum radius at $z=18$, the ring-like
structure creates a single-charged metastable vortex which
breaks-up subsequently into four fundamental solitons.}
\end{figure}

To demonstrate this phenomenon, we propagate numerically the
ring-shaped arrays of initially well separated coherently
interacting fundamental solitons in a saturable nonlinear medium.
\Fref{fig63} shows a characteristic example of a set of six
solitons which have their relative phases growing in steps of
$\vartheta =\pi/3$ along the ring, being initially directed to
collide with each other. This initial condition corresponds to the
inversion of the instability-induced ring break-up, so that the
solitons move towards the ring instead of flying away. We observe
highly inelastic collision of the solitons when they strongly
interact overlapping and loosing their identity. Nevertheless, the
initial phases of the solitons are tilted in such a way that the
total phase of the beams forms a screw dislocation in the ring
origin which prevents a simple fusion of all solitons. Instead,
the ring-shaped structure is formed. Similar situation is observed
in \fref{fig84} with an array of eight solitons and the formation
of a metastable vortex ring. Due to large amplitude modulations,
these intermediate (or metastable) structures never form ideal
stationary states, and they quickly split off to a set of new
isolated solitons, with the total number of splitters predefined
by the initial conditions. In this way, we were able to produce,
as a final state, the patterns with different number of solitons
by changing initial parameters, including the number and phase
tilt of solitons. Generally, the final number of solitons is
determined by the ring instability mode with a largest growth
rate, and in a saturable medium it is usually twice the
topological charge~\cite{Skryabin}. At the same time, we were able
to force single-charged `meta-vortex' in figures \ref{fig63} and
\ref{fig84} to split to three and four fundamental solitons,
respectively. The intermediate meta-state shows complex dynamics
of the instability development which may continue for several
tenths of diffraction lengths. Thus, the whole picture of the
soliton collision, ring formation, and the ring splitting is
somewhat similar to the `delayed-action interaction' recently
reported for interacting composite solitons carrying nonzero
angular momentum~\cite{action}. We note also that, in addition to
the known transformation of `spin-to-orbital' angular momentum
\cite{Skryabin}, in figures \ref{fig63} and \ref{fig84} we observe
a kind of `orbital-spin-orbital' transformation. The change of the
field momentum shown in \fref{fig63} is about $10\%$ of the
initial value, and it occurs only at the break-up stage.

\begin{figure}
\begin{center}
\epsfbox{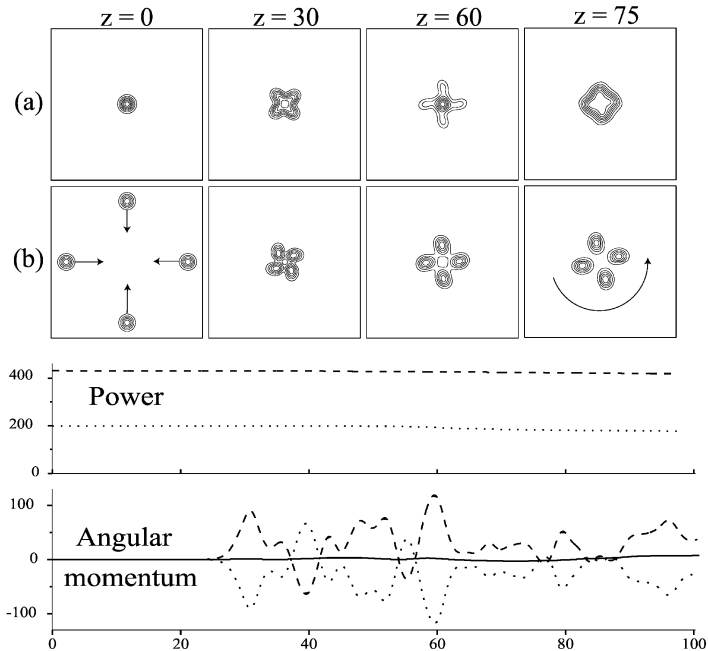}
\end{center}
\caption{\label{fig4v} Excitation of a four-lobe rotating vector
cluster, consisting of fundamental [row (a) and dotted lines] and
modulated [row (b) and dashed lines] mutually incoherent
components. Solid line: the total angular momentum. Arrows in the
row (b) show the initial transverse velocities of interacting
`atoms' and rotation of the final `molecule'.}
\end{figure}

Highly unstable ring-shaped beams are formed as a result of mutual
trapping and inelastic interaction of coherent solitons, as shown
in figures \ref{fig63} and \ref{fig84}. They represent the
intermediate steps in the process of the nonlinear switching
between the states with a different number of solitons. In the
context of the soliton algebra and all-optical switching, it might
be of interest to study in more details and to determine the
quantitative parameters necessary to obtain the final state with a
given number of solitons. At the same time, the mutual trapping of
coherent solitons is found to be too sensitive to the initial
perturbations in order to produce the metastable clusters.

Nevertheless, it is indeed possible to generate higher-order
optical beams, i.e. to synthesize the soliton molecules, from
interacting fundamental {\em coherent and incoherent} solitons. In
\fref{fig4v} we show the excitation of a long-lived four-lobe
vector cluster. We note that the four solitons in the
\fref{fig4v}(b) at $z=0$ are directed exactly to the center so
that there is no azimuthal tilt of their phase. As within the
cluster in \fref{fig1}, each soliton has a phase growing by a jump
$\vartheta=\pi/2$ along the ring, and this initial phase
distribution guarantees the appearance of the screw phase
dislocation, even the corresponding value of the total angular
momentum is very small (a solid line in \fref{fig4v}). Because
only the total angular momentum is conserved, but not the partial
momenta of the components, there is a freedom for components to
symmetrically exchange the angular momentum during the beam
propagation. After mutual trapping of all solitons at $z\approx
25$ the new-born vector cluster experiences strong radial
oscillations between the states with maximal (e.g. at $z=30$ and
$z=60$) and minimal partial angular momenta in the components, see
the diagram in \fref{fig4v}. At the same time, the $\pi/2$-phase
jumps introduced initially between the solitons (edge
dislocations) survive these strong oscillations, and the whole
cluster preserves its structure for a distance exceeding 100
diffraction lengths. Therefore, we observe the formation of a
composite state, the soliton molecule, by colliding simple
solitons with a nontrivial phase pattern.

\section{Conclusions}

We have studied the scattering and mutual trapping of several
fundamental solitons and the generation of soliton clusters and
vortex solitons -- the complex self-trapped states of light carrying
phase dislocations in the wave front. Inelastic collision of
solitons has been shown to result in the formation of ring-shaped
beams or metastable vortices which subsequently break-up creating
different number of fundamental solitons defined by the initial
conditions. This kind of the soliton delayed-action interaction
and the nonlinear transformation of the number of fundamental
solitons has been analyzed in the context of the soliton algebra
and all-optical soliton switching. We have also shown that the
vectorial interaction between the field components provides an
additional mechanism of the soliton cluster stabilization. In
particular, we have demonstrated the formation of vector soliton
clusters from colliding solitons, the process which can be
regarded as a synthesis of `soliton molecules'.

\ack A support from the Alexander-von-Humboldt Foundation and the
Australian Research Council is acknowledged.

\end{document}